\documentclass[aps,twocolumn,amsmath,amssymb,showpacs,prb]{revtex4}
\usepackage{epsfig}
\begin{document}
\begin{abstract}
We present a detailed NMR study of simple cubic CsC$_{60}$, the
only metallic cubic fullerides known so far besides
A$_{3}$C$_{60}$. $^{133}$Cs NMR signals the presence of about 12
to 15~\% of ``anomalous'' C$_{60}$ balls, characterized by a 15
meV gap. We present different experimental observations supporting
the idea that a spin-singlet (i.e. a C$_{60}^{2-}$) is localized
on these latter balls and stabilized by a JTD. A splitting of the
$^{133}$Cs spectrum into three different lines, with distinct
electronic environments, is observed which emerges naturally from
such a situation. Quadrupole effects on $^{133}$Cs confirm an
inhomogeneous distribution of the charge between C$_{60}$
molecules. Analysis of the relative intensities of the $^{133}$Cs
lines show that the spin-singlets do not form clusters at the
local scale, but are diluted within the lattice. This is probably
to avoid the occurrence of neighboring C$_{60}^{2-}$ and minimize
electrostatic repulsion between these balls. Through spin-lattice
relaxation measurements, we detect chemical exchange between the
Cs sites above 100 K. From this, we deduce that the lifetime of a
spin-singlet on a given ball decreases exponentially with
increasing temperature (from 15~sec at 100~K to 3 ms at 130~K).
The implications of the presence of such spin-singlets for the
nature of the metallic state is discussed, in relation with a
decrease of 1/T$_1$T for $^{13}$C at low temperatures, which would
be anomalous for a simple metal.
\end{abstract}

\title{Coexistence of spin-singlets and metallic behavior in simple cubic CsC$_{60}$}

\author{V. Brouet, H. Alloul }

\affiliation{Laboratoire de Physique des Solides, Universite Paris-Sud, Bat 510
91405 Orsay (France)
}

\author{L. Forr\'o}
\affiliation{
Laboratoire des solides semicristallins, IGA-Departement de
Physique, Ecole Polytechnique Federale de Lausanne, 1015 Lausanne
(switzerland)}

\date{\today} \newpage
\maketitle

\section{Introduction}

To investigate the electronic properties of fullerides, one would wish to
vary freely the number of doped electrons between 0 and 6 in the triply
degenerate t$_{1u}$ band. It is rather difficult with chemical doping
because only a limited number of phases appear to be stable. The FET devices
recently synthesized \cite{Batlogg} might offer a more convenient way to do
this, as the number of charge carriers injected in one C$_{60}$ monolayer
can be controlled through an applied gate voltage. However, although
transport measurements are easily accessible, all investigations cannot be
performed on these systems. Furthermore, it is not yet clear whether the properties of ``bulk C$%
_{60}$'' are similar to those of its surface \cite{Hesper2000} and a careful
comparison between the two systems is extremely desirable. To achieve that,
the number of chemically stable bulk stoichiometries has to be extended.

The conventional chemical doping is obtained by inserting $n$ alkali ions ($%
A $) in the C$_{60}$ structure, leading to A$_n$C$_{60}$
compounds, where the charge state of the C$_{60}$ molecule is very
close to $-n$. The direct influence of the alkali atoms on these
properties are thought to be negligible, although it could be a
complication in some cases, mainly through structural
modifications. Usually, only phases with an integer number of
electrons per ball are formed. A$_3$C$_{60}$ and A$_4$C$_{60}$
were discovered first \cite{Haddon} and were consequently the most studied. A%
$_3$C$_{60}$ are metals and superconductors at low temperatures, while A$_4$C%
$_{60}$ are insulators. As both should be metals in a simple band
structure approach \cite{Erwin}, such a difference raises
fundamental questions. Electronic correlation and/or electron
phonon coupling, both neglected in simple band calculations, must
be important here. In view of the contrasted experimental
situation, it seems important to extend investigations to other
stoichiometries to estimate the relative importance of the
different parameters. This paper is a second of a series of three
papers devoted to this task, called hereafter I
\cite{BrouetPart1}, II and III \cite{BrouetPart3} .\medskip

    Certain stoichiometries are more difficult to obtain
for the following reasons.\\
 - n=1 is stable (with A=K, Rb, Cs) but
spontaneously polymerizes below 350 K \cite{ChauvetPRL94}, leading
to a rich but different physics, where low dimensionality could
play a role. To avoid this problem, the cubic $fcc$ phase (T $>$
350 K) has been quenched to obtain metastable phases with cubic
symmetry. A non-cubic phase is usually formed with C$_{60}$ dimers
\cite {Dimeres}. In the case of CsC$_{60}$ however, a cubic phase
can be obtained by quenching to liquid nitrogen
\cite{KosakaPRB95}, which is orientationally ordered ($sc$
Pa$\overline{3}$ symmetry) \cite{LappasJACS95}. It is called
hereafter CQ for cubic quenched and the purpose of this paper is
to study the electronic properties of this phase, which is
predominantly metallic. On the other hand, the properties of the
high temperature cubic phase will be discussed in paper III.

- n=2 is not obtained with large alkali ions (K, Rb, Cs) that
first fill the large octahedral site (forming A$_1$C$_{60})$ and
then the two tetrahedral sites (forming A$_3$C$_{60})$.
Na$_2$C$_{60}$ is the only case where n=2 can be studied and we
have shown that it behaves similarly to A$_4$C$_{60},$ i.e. there
is a gap in the electronic structure \cite{BrouetPRL2001}. As
these stoichiometries have symmetric positions in the t$_{1u}$
band, this gap must be an intrinsic feature of these band
fillings. A detailed comparative NMR study of Na$_2$C$_{60}$ and
K$_4$C$_{60}$ has been presented in paper I.

- n=5 cannot be reached by inserting only alkali ions. A charge
transfer of 5 electrons per C$_{60}$ has been observed in
ABa$_2$C$_{60}$ (A=K, Rb, Cs) because Ba gives 2 electrons to the
C$_{60}$ \cite{PRBYldirim96}. The t$_{1u}$ band might be modified
in this case by some hybridization with Ba orbitals. We note that
the first investigations indicate a metallic state, a Korringa law
has for example been observed by NMR \cite{PRBThier99}, although
more studies are still necessary. \medskip

This short review suggests that, beyond the opposition of A$_3$C$_{60}$ and A%
$_4$C$_{60}$, compounds with an \textit{even} number of electron
per C$_{60}$ are insulators, while those with an \textit{odd}
number of electrons are metals. This is puzzling as, if the band
picture fails because the Coulomb repulsion U~$\approx $~1 eV is
larger than the bandwidth W~$\approx $~0.5 eV, but, all compounds
with an integer number of electrons should be Mott insulators. The
study presented here reveals the presence in CQ CsC$_{60}$ of a
small number of C$_{60}^{2-}$. This implies that there must be an
effective attractive interaction at the local scale that helps in
this case to overcome the strong Coulomb repulsion. Understanding
the actual mechanism for the formation of such pairs is then
likely to shed new light on how some of these phases can become
metallic despite the large Coulomb repulsion. We will argue that a
Jahn-Teller distortion (JTD) of the C$_{60}$ molecule, which is a
consequence of the electron-phonon coupling in these systems with
degenerate band, is the missing ingredient responsible for this
behavior.

\begin{figure}[t]
\centerline{ \epsfxsize=0.4 \textwidth{\epsfbox{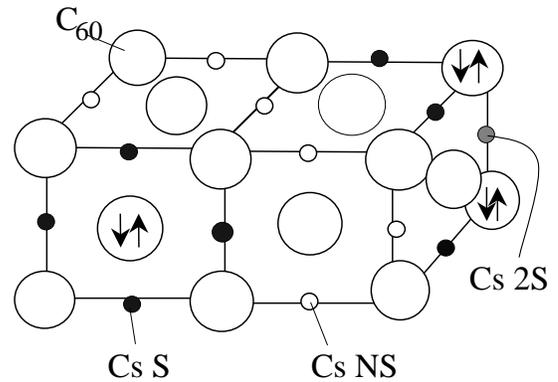}} }
\caption{Structural model of cubic quenched CsC$_{60}$ describing
the electronic properties as revealed by NMR. Spin-singlets are
localized on a fraction of the C$_{60}$ balls (about 10\%) leading
to different types of Cs sites. They are called respectively NS, S
or 2S when 0, 1 or 2 of the 6 C$_{60}$ neighbors of one Cs site
bears a singlet. The remaining electrons are delocalized over the
other C$_{60}$ balls.} \label{model}
\end{figure}

The difficulties introduced by the quench necessary to produce CQ CsC$_{60}$
have limited so far experimental investigations, for example there are no
reported transport measurements. Also, the available temperature range is
restricted to T$<135$ K, above which it transforms irreversibly into the
dimer phase. The CQ phase appears to be metallic as ESR detects a Pauli-like
susceptibility \cite{KosakaPRB95} and $^{13}$C NMR spin-lattice relaxation a
Korringa law down to 50 K \cite{BrouetPRL99}. By analogy to A$_3$C$_{60},$ a
superconducting ground state could be expected, but no superconductivity has
been detected down to 4 K \cite{KosakaPRB95}. We have suggested in ref. \cite
{BrouetPRL99} that it could be due to a competition with alternative ground
states. Indeed, clear anomalies in the $^{133}$Cs NMR spectrum indicate that%
\textit{\ the electronic properties of CQ CsC}$_{60}$%
\textit{\ are not homogeneous on the local scale}. The situation
can be summarized by the picture of Fig. \ref{model}. Within a
predominantly metallic phase, we proposed that 2 electrons get
localized on some C$_{60}$ balls (about 10 \%) and paired into a
spin-singlet. We will give in this paper new experimental
observations which support this scenario (section II). To get more
detailed information on this charge segregation phenomenon, we
particularly focus our attention in this paper on some features
that were not addressed in our
previous report on this phase \cite{BrouetPRL99} : the distribution of the C$%
_{60}^{2-}$ within the metal (section III) and their lifetime
(section IV). Finally, we discuss the implications of the presence
of C$_{60}^{2n-}$ for the nature of the metallic state (section
V).

\section{Three Cs lines}

\begin{figure}[tbp]
\centerline{ \epsfxsize=0.42 \textwidth{\epsfbox{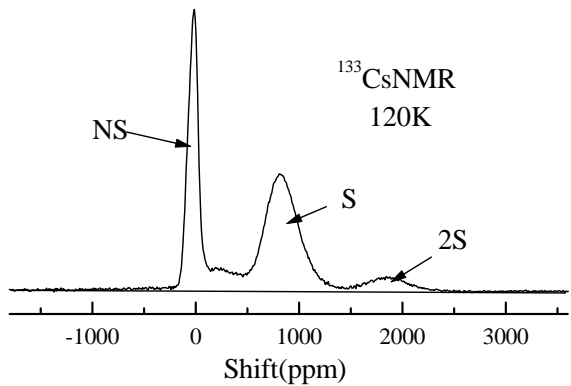}} }
\caption{$^{133}$Cs NMR spectrum at 120 K in the cubic quenched
phase showing the different Cs lines. The repetition time was 200
msec. } \label{2S}
\end{figure}

On the $^{133}$Cs spectrum of Fig. \ref{2S}, obtained at 120 K and
7 T, three different lines are clearly resolved at -15~ppm, 800
ppm and 1800 ppm. The latter line has a much smaller intensity (6
$\pm 2\%$ of the total) than the two main lines and was not
detected in ref \cite{BrouetPRL99}. As shown hereafter, its
detection allows to confirm and refine the model of Fig. \ref
{model}.

Finding three Cs lines contradicts the expectations of structural
studies, for which there is only one site for Cs in this phase,
namely the octahedral site \cite{LappasJACS95}. With one site,
only one NMR line should be observed, as in the high temperature
$fcc$ phase \cite{TyckoPRB93}. The particular phase diagram of
CsC$_{60}$ allows to convince oneself rather directly that all
these lines are intrinsic, because they all disappear irreversibly
at the transition to the dimerized structure. In addition, SEDOR
experiments reported in ref. \cite{BrouetPRL99} demonstrate that
the different Cs sites are mixed on the microscopic scale.

\subsection{Local electronic environment of the three Cs lines}

\begin{figure}[t]
\centerline{ \epsfxsize=0.45 \textwidth{\epsfbox{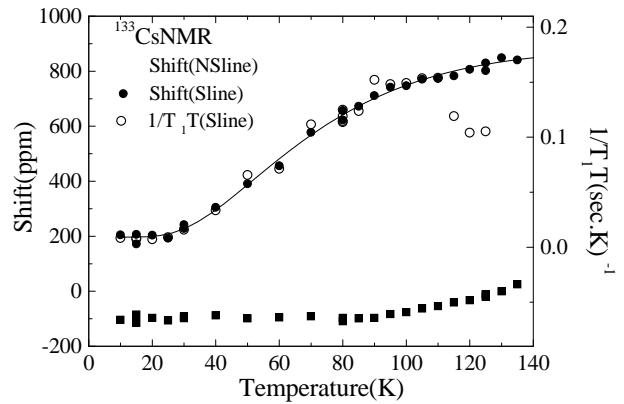}} }
\caption{$^{133}$Cs NMR shifts with respect to a CsCl solution as
a function of temperature for the two main Cs lines (NS and S) in
cubic quenched CsC$_{60}$. For the S line, 1/T$_{1}$T is also
displayed by open symbols (right scale). The line is a fit to a
law A+B/T*exp(-Ea/$k_B$T) with Ea=15 meV.} \label{Csshift}
\end{figure}

The splitting of the $^{133}$Cs NMR spectrum implies that there
are three different Cs sites well defined at the local scale.
 To understand their origin, the study of the local
electronic environment, as probed by the NMR shift $K$ and the spin-lattice
relaxation rate $1/T_1$ of each line, both presented in Fig. \ref{Csshift}%
, is very helpful. These quantities depend on the hyperfine coupling $A$
between Cs and one of its 6 C$_{60}$ first neighbors.
The shift is expected to read $K=\sigma +6A\chi _{loc}$, where $\sigma $ is
the reference chemical shift and $\chi _{loc}$ the local electronic
susceptibility, and $1/T_1\propto 6A^2\chi _{loc}^{\prime \prime }(\omega
_0)/\omega _0,$ where $\chi _{loc}^{\prime \prime }(\omega _0)$ is the
imaginary part of $\chi _{loc}$ at the nuclear Larmor frequency (for $^{133}$%
Cs, $\omega _0=40$ MHz at 7 T). The reference chemical shift can
be estimated by the position of Cs$^{+}$ in the gas phase to be
around $\sigma =$-300 ppm \cite{Slichter}. Therefore, the shift of
the less shifted line is small, about 200 ppm at low temperature,
and we label this line as NS for Not Shifted. Such a small shift
is expected for Cs in metallic fullerides because electrons are
almost completely transferred to the C$_{60}$ balls. In addition,
it is almost temperature independent according with a metallic
susceptibility (the upturn above 100 K will be discussed in
section IVb). On the contrary, the two other lines exhibit
unusually large shifts with large temperature dependences
inconsistent with a metallic environment. To contrast with the NS
line, the line at 800~ppm is called S (for shifted).
Interestingly, the shift of the most shifted line is approximately
twice that of the S line and we therefore label it 2S.\ Similarly
its T$_1$ is reduced by a factor 2 with respect to that of the S
line.\ At 80 K, $^{2S}$T$_1$=~43ms~$\pm $ 11ms to be compared with
$^{S}$T$_1$~=~110 $\pm $ 10 ms for the S line.

These latter points strongly suggests that
\textit{the inequivalency between
the Cs sites results from a local perturbation and that the three lines
emerge from the coupling with zero (NS), one (S) or\ two (2S) ``anomalous C}$%
_{60}$\textit{''}. Indeed, if we call $\chi _g$ and $\chi _m$ the
susceptibilities of respectively one ``anomalous C$_{60}$'' and
one ``metallic C$_{60}$'', A$_g$ and A$_m$ the hyperfine couplings
between them and Cs, we expect the following relations.

\begin{eqnarray*}
K_{NS} &=&6A_{m}\chi _{m}\text{ \ \ \ \ \ \ \ \ \ \ \ \ \ \ \ \ }%
1/T_{1})_{NS}\propto 6A_{m}^{2}\chi _{m}^{\prime \prime } \\
K_{S} &=&5A_{m}\chi _{m}+A_{g}\chi _{g}\text{ \ \ \ }%
1/T_{1})_{S}\propto 5A_{m}^{2}\chi _{m}^{\prime \prime }+A_{g}^{2}\chi
_{g}^{\prime \prime } \\
K_{2S} &=&4A_{m}\chi _{m}+2A_{g}\chi _{g}\text{  \ }%
1/T_{1})_{2S}\propto 4A_{m}^{2}\chi _{m}^{\prime \prime }+2A_{g}^{2}\chi
_{g}^{\prime \prime }
\end{eqnarray*}
\qquad

This fits the experimental observations $K_{2S}\approx 2K_S$ and $%
1/T_1)_{2S}\approx 2$ $1/T_1)_S$ provided that the anomalous
component dominates the metallic one, i.e $A_g$ $\chi _g>>$ $A_m$
$\chi _m,$ which is obvious since $K_S>>K_{NS}.$ We can actually
estimate $A_m$~$\chi _m$=~30~ppm and $A_g$~$\chi _g$~=~700~ppm at
120 K.

Fig. \ref{Csshift} also shows that there is a scaling between K and 1/T$_1$%
T for the S line up to 110 K, which ensures that they are both
dominated in this temperature range by the same electronic
excitations. This common behavior represents $\chi _g$ and is
fitted on Fig. \ref{Csshift} by an activated law
(A/T)*exp(-E$_a$/k$_B$T) with E$_a\approx $15 meV. This means that
the ``anomalous C$_{60}"$ correspond to a molecular arrangement
with a gap E$_a$ and therefore electrons must be localized on
these balls. The simplest idea would be that one electron is
``trapped'' by some defects, but in this case we should observe a
paramagnetic behavior of this localized spin 1/2 at low
temperatures, which we do not. The most plausible way to
explain the vanishing susceptibility at low $T$ is that \textit{two }%
electrons are localized instead of one and paired up into a \textit{singlet}%
. Molecular calculations \cite{Manini} indeed predicts that the
singlet state wins over the triplet state (favored by Hund's
rules) for C$_{60}^{2-}$ in the presence of a JTD.\ In our
opinion, this gain of energy associated with JTD is at the origin
of the formation of C$_{60}^{2-}$ within the metal. We note that
there might be an additional temperature independent contribution
$\chi _0$ to K as its low temperature value is very different from
our estimation of $\sigma =-300$ ppm or from the shift of the NS
line.

Na$_2$C$_{60}$ is a natural reference for the NMR behavior of the JTD C$%
_{60}^{2-}$ \cite{BrouetPRL2001}. The gap measured on Cs S sites
is much smaller than the 140 meV singlet-triplet gap of a
C$_{60}^{2-}$ measured through 1/T$_1$ relaxation in
Na$_2$C$_{60}.$ The limited stability of the CQ phase in
temperature (T $<$ 135 K) does not really allow to probe the
existence of molecular excitations of this high energy. This
15~meV gap more likely corresponds to excitations of the singlets
towards the metallic band. We have seen in paper I that in
Na$_2$C$_{60}$ also, the low temperature behavior is governed by
such band excitations. The fact that the spin-singlets cannot be
viewed as an isolated entity might also explain why $K$ and
$1/T_1$ seem to tend to a constant value different from zero at
low temperatures. More details will be given in the last section
of this paper on the interactions between the singlets and the
band.

\subsection{Confirmation of local charge segregation}

This inhomogeneous distribution of the charge among the C$_{60}$ balls will
create electric field gradients (EFG) at the Cs S and 2S sites and since $%
^{133}$Cs is a spin 7/2, it should sense them through its
quadrupole moment Q. Actually, we show hereafter that the larger
linewidth of the S line ($\Delta \nu \approx 16$~$kHz$ at half
width) compared to the NS line ($\Delta \nu \approx 5$~$kHz$) is
due to larger quadrupole effects on these sites.

In
presence of quadrupole effects, the different nuclear transition (7/2$%
\leftrightarrow 5/2,$ 5/2$\leftrightarrow 3/2,$ etc) acquire
different frequencies $\nu _0\pm n$ $\nu _Q$ $f(\theta ,\varphi
)$, where n is the order of the transition, $\nu _Q$ a quadrupole
frequency proportional to
the strength of the EFG and the value of the quadrupole moment, and $%
f(\theta ,\varphi )$ a function depending on the orientation of
the EFG with respect to the NMR field. Eventually, if $\nu _Q$ is
very large, the ``satellite transitions'' could be lost because
they become very broad and only the central transition
1/2$\leftrightarrow -1/2$ \ is observed. This is not very likely
for $^{133}$Cs, because it has one of the smallest quadrupole
moment of all nuclei. The lengths of the NMR pulses that optimize
the signal should be different in this situation \cite{pulse}. As
this is not the case, we conclude that all the nuclear transitions
are observed for all sites and the broadening of the spectrum is
then a first order quadrupole broadening.

\begin{figure}[tbp]
\centerline{ \epsfxsize=0.45 \textwidth{\epsfbox{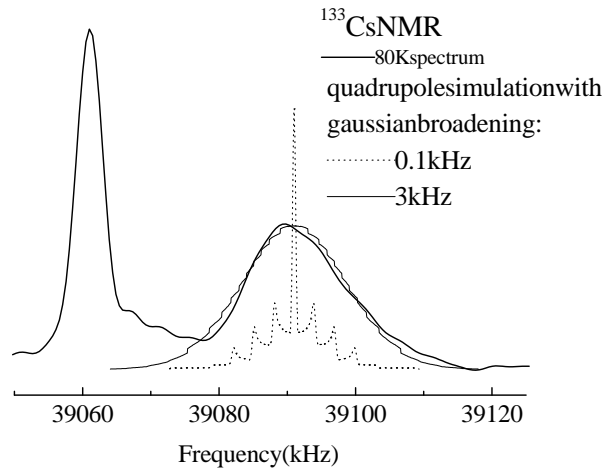}} }
\caption{$^{133}$Cs spectrum at 80 K in the CQ phase (thick line).
The other lines are quadrupole simulations of the S lineshape
explained in the text. The theoretical lineshape is convoluted by
a gaussian of linewidth 0.1 kHz (dotted line) and 3 kHz (thin
line).} \label{Quad}
\end{figure}

The value of $\nu _Q$ expected for the $S$ sites in our model can be
estimated by a direct calculations of the EFG, assuming point charges.\
$$
\nu _Q=\frac{3(1-\gamma _\infty )eV_{zz}Q}{2I(2I-1)h}\text{ \ \ \ with \ \ }%
V_{zz}=\frac{\partial ^2V}{\partial z^2}
$$
where $V$ is the electrostatic potential, $I$ the nuclear spin and $%
(1-\gamma _\infty )$ the Sternheimer antishielding factor
\cite{Slichter}. The distribution of charges around one S site
yields $\nu _Q\approx 6$~$kHz$. The resulting lineshape depends
also on the other principal values of the EFG, here we calculate
$V_{xx}=V_{yy}=-1/2$ $V_{zz}$. A simulation of the spectra
(for this powder sample, it has to be averaged on all possible orientations  \cite{Slichter}%
) is shown on Fig.~\ref{Quad} in two different cases. For the
dotted line, a very small experimental broadening is introduced
and all satellite transitions are visible. For the thin line, a
realistic experimental broadening of 3 kHz, similar to the NS
linewidth, is introduced, which allows a direct comparison with
the experimental spectrum. The agreement is very good which should
only be taken as a proof of the consistency of our analysis. It is
known that the EFG could vary significantly from this rough point
charge estimate, when a more realistic situation is considered,
for example an inhomogeneous distribution of the charge over the
C$_{60}$ molecule.

As for the 2S line, the linewidth is similar to that of S ($\Delta
\nu \approx 18$ $kHz$). This is consistent with the idea that the
broadening is not of magnetic origin, for example arising from a
distribution of shift, in which case the linewidth should double.
The quadrupole broadening for a 2S site in our model depends on
the relative position of the C$_{60}^{2-}$ with respect to this Cs
site. If they are on a straight line with respect to their common
Cs neighbor, the point charge calculation predicts that $\nu
_Q[2S]=2$ $\nu _Q[S],$ while if they are in different directions
$\nu _Q[2S]\simeq \nu _Q[S]$. This latter case would apparently be
in better agreement with the data but, as already mentioned, the
calculation of the EFG is somewhat rough and the lack of structure
in the S and 2S lineshape does not allow a refined analysis.
Actually, next section provides more reliable information on the
distribution of the C$_{60}^{2-}$, which do not favor this
arrangement.

The detection of quadrupole effects on S and 2S sites is however
essential to ensure a qualitative agreement with the model of Fig.
\ref{model}.

\section{Distribution of the C$_{60}^{2-}$ within the metal}

\begin{figure}[b]
\centerline{ \epsfxsize=0.45 \textwidth{\epsfbox{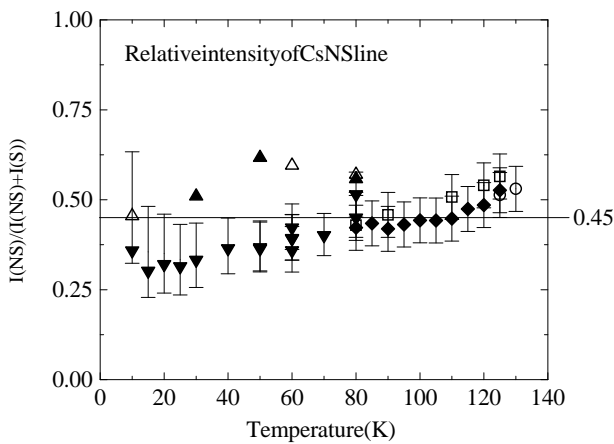}} }
\caption{Relative intensity of the $^{133}$Cs NS line compared to
the one of the S line. This gives the percentage of NS sites as a
function of temperature. The different symbols refer to different
samples and/or different quench, showing a good reproducibility of
this measurement. Error bars are larger at low temperatures when
the two lines begin to overlap. } \label{intensite}
\end{figure}

Following the lines of the model described in Fig. \ref{model},
the number of C$_{60}^{2-}$ can be deduced from the relative
intensities of the three lines, the area of one line being
proportional to the number of sites resonating at this frequency.
As can be seen on Fig. \ref{intensite}, the ratio between S and NS
sites is nearly constant as a function of temperature,
NS/S~=~45/55, and it is reproducible with different samples and/or
different quench, so that it must correspond to a well defined
equilibrium. As there are 6 C$_{60}$ neighbors to one Cs site,
there must be roughly 10 \% of C$_{60}^{2-}$ to produce this
ratio. With such a number,
there is a non-negligible probability of finding a Cs site with two C$%
_{60}^{2-}$ neighbors, which correspond to our ``2S sites''.

\subsection{Structural arrangement of the C$_{60}^{2-}$}

\begin{figure}[tbp]
\centerline{ \epsfxsize=0.45 \textwidth{\epsfbox{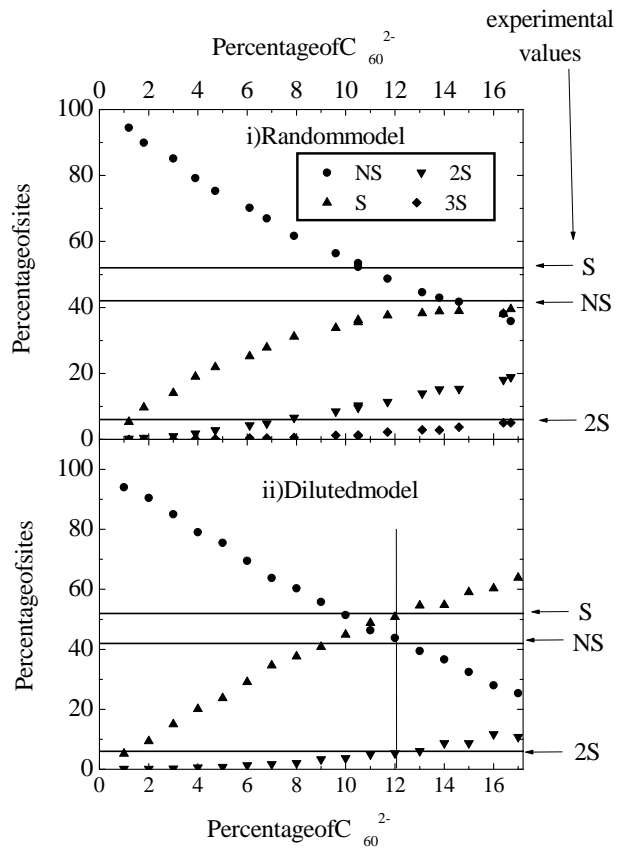}} }
\caption{Simulation of the number of NS, S, 2S and 3S Cs sites as
a function of the percentage of C$_{60}^{2-}$ in two different
situations: i)the distribution of C$_{60}^{2-}$ is assumed to be
random, ii) a ``diluted model'' where two neighboring C$_{60}$
molecules in a singlet state are excluded (in this case there are
no 3S sites). Experimental values at 120 K are given by horizontal
lines for comparison and the agreement with 12 \% C$_{60}^{2-}$ in
the diluted case by a vertical line.} \label{distribution}
\end{figure}

More precisely, a closer look at the relative intensities, and
especially the number of 2S sites, can help us to address the
question of the \textit{distribution} of C$_{60}^{2-}$ balls
within the metal. On Fig. \ref{distribution}, we show the results
of calculations of the number of Cs sites as a function of the
concentration of C$_{60}^{2-}\,$in two different cases :

i) with a random distribution of C$_{60}^{2-}$

ii) when neighboring C$_{60}^{2-}$ are excluded (``diluted" case).

To compare with the experiment, the number of S, NS and 2S sites
at 120K are reported on Fig. \ref{distribution}. In the random
case, the calculated intensities never quite reach the
experimentally observed values. The fraction of 2S sites is
rapidly higher than in the experiment and there are always more NS
sites than S sites, contrary to the experiment. Furthermore, the
intensity of a 3S line corresponding to C$_{60}$ with three
C$_{60}^{2-}$ neighbors would be sizable, although it is not
observed. In the diluted case, 12\% of C$_{60}^{2-}$ almost
exactly correspond to the experiment at 120 K, as shown by the
vertical line.$\;$In this situation, the 3S configuration is
forbidden, according to the experimental finding. Therefore, we
conclude that \textit{C}$_{60}^{2-}$ \textit{are diluted within
the lattice to avoid a situation where they are first neighbors}.
They do not segregate in superstructures that could be reminiscent
of the formation of stripes. This last point is consistent with
SEDOR experiments that rule out the formation of ``clusters'' of
C$_{60}^{2-}$. This dilution could be favored because it minimizes
the electrostatic repulsion between C$_{60}^{2-}$. As a function
of decreasing temperature, comparison between Fig.~\ref{intensite}
and Fig.~\ref{distribution}
suggests an increase of the number of C%
$_{60}^{2-}$ to about 15 \%, although the error bars are quite large.\medskip\

\subsection{What limits the number of C$_{60}^{2-}$ to about 10 to 15\% ?}

The experimental concentration of C$_{60}^{2-}$ appears a little
bit puzzling at first, since one could expect that if the
C$_{60}^{2-}$ are more stable than C$_{60}^{-},$ there should be a
total charge dismutation between C$_{60}$ and C$_{60}^{2-}.$ One
problem with such a dismutated ground state is that it
\textit{cannot} be realized in this phase because of the inherent
frustration of the $fcc$ lattice. Furthermore, if first
neighboring C$_{60}^{2-}$ are excluded, the maximum number of
C$_{60}^{2-}$ would already be reduced to 25 \%. This  would
correspond to a completely ordered structure with a singlet on
each corner of the cubic cell and none on the faces. With some
disorder, this number would rapidly drop.

Maybe the C$_{60}^{2-}$ do not order collectively, because they
are preferentially formed on some defects of the structure. A
particular feature of the orientational order characterizing the
$sc$ phase is that there are two orientations for the C$_{60}$
molecule. They are defined by an angle $\varphi$ from which they
are rotated with respect to the so-called ``standard orientation"
around one diagonal
axis \cite{Launois}. The value minimizing the C$%
_{60}$\ interactions occurs for $\varphi =98^{\circ }$\, but a second minimum for $%
\varphi =38^{\circ }$\ is nearly degenerate. The fit
of the X-ray spectra can be improved in all $sc$ phases (C$_{60}$, Na$_2$AC$%
_{60}$\ and CQ CsC$_{60}$) by allowing 12 to 20 \% of the balls at
low T to be in a ``minor orientation" corresponding to $\varphi
=38^{\circ }$. For CQ CsC$_{60}$\ 16 \% gives the best fit at 4.5
K \cite{LappasJACS95}. The closeness of this value with the number
of singlets found here is such that one is tempted to consider
that a ball in a
minor orientation might be viewed as a disorder potential efficient in trapping a C$%
_{60}^{2-}$. It is indeed known that the overlap integrals between
neighboring C$_{60}$\ are very sensitive to their relative
orientations. In pure C$_{60}$, no structural correlations were
found between the balls in a given orientation \cite{Launois}, as
found here for the singlets.\

\section{Motion of spin-singlets above 100~K}

The observation of three different lines indicate that the spin-singlets are
nearly static on well determined C$_{60}$\ balls on the NMR timescale.
However we show hereafter that other experimental observations such as the
NMR spin lattice relaxation time T$_1$\ allow to determine the lifetime of
the spin singlets which decreases with increasing temperature.

\subsection{Relaxation behavior}

\begin{figure}[t]
\centerline{ \epsfxsize=0.45 \textwidth{\epsfbox{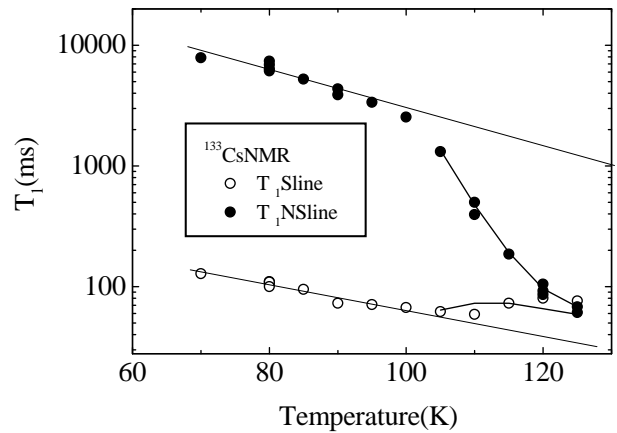}} }
\caption{T$_{1}$ for the two main $^{133}$Cs NMR lines between
80~K and 135~K. The reduction of the NS line T$_{1}$ for  above
100~K is attributed to chemical exchange between the two sites.
The straight lines extrapolate the temperature evolution of
T$_{1}$ if there were no exchange. The other lines are a fit to
our model of chemical exchange (see text).}\label{T1a}
\end{figure}

As can be seen in Fig. \ref{T1a}, T$_1$ for the NS line changes
dramatically between 110 K and 130 K, getting shorter by two order
of magnitudes. At 130 K although the two main $^{133}$Cs lines are
perfectly resolved, they have almost the same T$_1$. This finding
suggests that the nature (NS, S or 2S) of any Cs nucleus is
changing with time, so that the properties of these sites become
identical on the large timescale of the T$_1 $\ measurement (more
than 50 ms), although on the short timescale corresponding to the
inverse spectrum width (typically 100~$\mu $s here), individual
resonance frequencies are still resolved. Within our model, such a
situation would occur if the electronic spin-singlets start to
jump from a C$_{60}\;$to its near neighbor C$_{60}$. When such an
electronic jump occurs, some\ Cs S sites become NS and vice-versa.

\begin{figure}[b]
\centerline{ \epsfxsize=0.35 \textwidth{\epsfbox{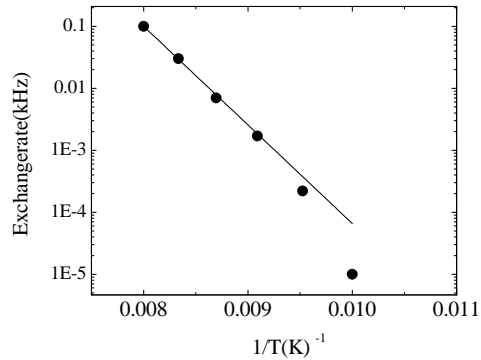}} }
\caption{Logarithmic plot of the value of the exchange parameter
$p$ used to obtain the line of Fig. 7 as a function of the inverse
of the temperature. $p$ can be fitted by an activated law as shown
by the solid line.} \label{T1b}
\end{figure}

Let us examine in more details whether a motion of the
spin-singlets could explain the change in T$_1$ \cite{Abragam}. We
assume that there is a probability $p$ per unit time that the
nature of the site changes from S to NS or NS to S. On the other
hand, we suppose that for each situation (Cs neighboring a
C$_{60}^{2-}$ or not), there is a well defined relaxation rate
W$_S$ (resp. W$_{NS}$). Let $n_S$ be the number of S sites (resp.
$n_{NS}$ for NS sites) excited at the time t=0 by the saturation
pulse of the T$_1$ experiment, the relaxation of the two spin
species obeys the following coupled system.

\begin{eqnarray*}
\frac{dn_{S}}{dt} &=&-(W_{S}+p)\text{ }n_{S}+p\text{ }n_{NS} \\
\frac{dn_{NS}}{dt} &=&p\text{ }n_{S}-(W_{NS}+p)\text{ }n_{NS}
\end{eqnarray*}

For an infinite lifetime of the singlets, that is for p = 0, we recover the
expected exponential decay with $(1/T_1)_S=W_S$ and $(1/T_1)_{NS}$ $~=$ $%
~W_{NS}$. When W$_{NS}\leq $ p $\leq $ W$_S,$ the relaxation rate
of the NS line is strongly reduced and tend to the value
(W$_S+$W$_{NS})$ $/$ $2$ as p increases. By extrapolating the
variation of W$_S$ and W$_{NS}$ by straight lines (see
Fig.\ref{T1a}), we calculate with this model the value needed for
$p$ to obtain the experimental T$_1$ values. We can indeed
reproduce the experimental results, as shown by the dashed lines
on Fig. \ref{T1a}. For each temperature, the value of $p$ that we
have used is reported on Fig. \ref{T1b}. It follows an activated
temperature dependence with p~=~5~10$^{14}*\exp (-3700/$T), which
sets a new energy scale for the sytem of about 320 meV,
characterizing the energy barriers that traps the singlet. This
magnitude is quite comparable with the activation energies usually
found for the molecular rotations in A$_n$C$_{60}$ compounds. The
temperature range where we start to observe the motion of the
spin-singlets also roughly corresponds to the usual range for the
onset of molecular motions. This reinforce the idea that the
trapping of the singlets could be related to C$_{60}$ in a minor
orientation.

From the frequency of jumps between
the two sites, we can deduce the characteristic lifetime of a spin-singlet
on a particular C$_{60}$ ball. It changes from 15 sec at 100 K to 3 ms at
130 K.\

\subsection{Static spectrum}

This characteristic time is still long compared to the time scale of the
experiment, therefore the motions of the singlets do not show up clearly on
the static spectra. However, in the range of temperatures for which the T$_1$%
\ become identical, the scaling between the shift and the
relaxation for the S line does not hold anymore (Fig.
\ref{Csshift}), which suggests that the jumps of the singlets
begin to affect as well the resonance frequencies.\ If we could
increase the temperature further, we could expect that p will
become short enough to yield a \textit{motional narrowing }of the
static spectrum with three lines ultimately merging into one. From
the data of Fig. \ref{T1b}, we can extrapolate that the two main
lines would merge together at 150 K, when the frequency jump is
comparable to the 30 kHz frequency separation of the two lines.\
This is a first step to reconcile the properties of CsC$_{60}$ in
its CQ and high temperature phases, where only one Cs line is
observed. The relation between these two cubic phases will be
discussed in more details in paper III. The upturn of the NS shift
above 110 K that can be noted on Fig. \ref{Csshift} might be a
precursor sign of this effect.

\section{Interplay between the spin-singlets and the metallic state}

The motion of spin-singlets described above shows that they can be
considered to some extent as charge carriers. Actually, in a
random metallic state with negligible on site Coulomb repulsion
 and nominally one electron per C$_{60}$, one expects to find on each C$_{60}$ either zero, two or
one electron, either with spin up or down, all states with equal
probabilities. A 25 \%\ concentration of doubly charged C$_{60}$,
or less if we include some Coulomb repulsion,
 is then quite natural and it is more their increased lifetime that is surprising here.
 We assign it to the higher stability of the JTD C$_{60}^{2n-}$,
 and this could play a particular role in the charge transport of A$_{2n+1}$C$_{60}$.  In the presence of any disorder potential,
 they could become easily trapped, which seems to be the case in CQ-CsC$_{60}$.

 An important question that has been left aside so far is
whether there is also 12 \%\ of static neutral C$_{60}$ or if the
unpaired electrons are delocalized on the 88\%\ remaining balls.
Although hypothetic, the possible ``non-stoichiometrie" of the CQ
phase is reminiscent of the discussions about the role of
vacancies in A$_{3}$C$_{60}$. The physical model behind this idea
was that A$_{3}$C$_{60}$ would be a Mott insulator for integer
filling because of the strong electronic correlations and that
only the deviation from 3, introduced by a reproducible number of
vacancies, allows the formation of a metal \cite{Lof}. The same
argument could be discussed for CQ CsC$_{60}$.

\begin{figure}[tbp]
\centerline{ \epsfxsize=0.45 \textwidth{\epsfbox{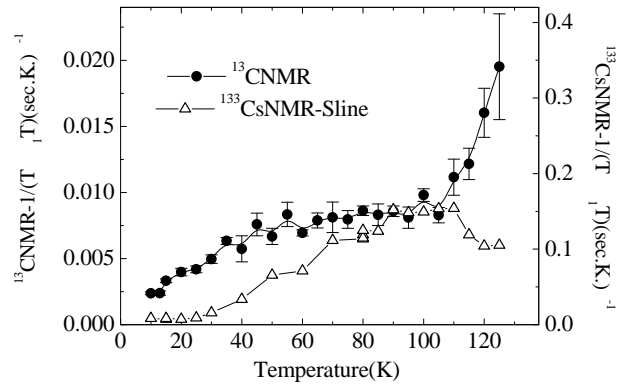}} }
\caption{1/T$_1$T as a function of temperature for $^{13}$%
C NMR (left scale) and $^{133}$Cs NMR S line (right scale) in the
CQ phase of CsC$_{60}$. While the $^{133}$Cs S line behavior
mainly reflects the properties of the gapped C$_{60}^{2-}$ ion,
$^{13}$C NMR is dominated by the metallic behavior which becomes
anomalous below 50 K.} \label{T1Tcar}
\end{figure}

From the experimental point of view, we are not able to
distinguish by $^{13}$C\ NMR\ the signals of differently charged
C$_{60}$ balls, neither the signal of C$_{60}^{2-}$ or of any
eventual neutral C$_{60}$, from that of the dominant fraction of
C$_{60}^{-}$. This results from the well known fact that the
anisotropic orbital contribution to the NMR shifts dominates the
electronic contribution to the shift in A$_{n}$C$_{60}$ compounds
and broaden the signals at low temperatures
\cite{reviewPennington}. Although different relaxation times could
be expected for the different components, it is beyond
experimental accuracy to detect such a small fraction of distinct
behaviors. The relaxation curves for the NMR magnetization, from
which T$_1$ is extracted, are intrinsically
 not exponential at low temperatures in
fullerides, because of the existence of inequivalent $^{13}$C
sites
 on a given molecule \cite{BrouetPart1}.
 This would easily
blur the occurrence of different relaxation mechanism for a small fraction
of the C$_{60}$.

Nevertheless, we do observe an anomaly in $^{13}$C NMR, which
could help us to understand the coupling between the band and the
formation of C$_{60}^{2-}$. Fig. \ref{T1Tcar} shows that although
1/T$_{1}$T is constant between 50~K and 100~K as expected for a
metallic Korringa law, it decreases below 50~K by a factor
$\approx 3$ (the upturn above 100~K will be discussed in paper
III). As emphasized on Fig. \ref{T1Tcar}, this decrease is not
correlated with the $^{133} $Cs NMR for which 1/T$_{1}$T saturates
precisely below 50 K and therefore more likely affects the
properties of the band that dominate anyway the $^{13}$C NMR. The
Korringa law relates 1/T$_{1}$T to the square of the density of
states (1/T$_{1}$T $\propto $ n(E$_{f})^{2}$) which suggests a
decrease in the density of states at low T, that could be due to
an increase in the number of C$_{60}^{2-},$ taking more electrons
out of the band. The general trend of the variation of the number
of C$_{60}^{2-}$ indicated by Fig. \ref{intensite} is in
qualitative agreement with this idea but not quantitatively. The
maximum variation of the number of NS sites between 50 K and 10 K
would be from 50 \% to 25 \%, corresponding to an increase in
C$_{60}^{2-}$ from 12\% to 17 \% (see Fig. \ref{distribution}b)
and
then a decrease of n(E$_{f}$) by about only 5 \%. If the decrease of 1/T$%
_{1} $T is related to this effect, the properties of the band must
be modified at the same time, for example 1/T$_{1}$T can be
enhanced above the value expected from n(E$_{f}$) by correlation
effects \cite{Slichter}. Actually this effect exists in CQ
CsC$_{60}$, as the constant value of 1/T$_{1}$T is enhanced by
almost a factor 2 compared to Na$_{2}$CsC$_{60}$ that has a
similar susceptibility and this could change as one goes further
away from integer filling.

\section{Conclusion}

We have shown that many independent NMR observations support the
idea that spin-singlets are present in the CQ phase of CsC$_{60}$
on about 12 \% of the C$_{60}$ balls. The lifetime of these
spin-singlets on a given ball increases exponentially with
decreasing temperature (from 3 ms at 130 K to 15 sec at 100 K). We
believe that they are formed randomly by hopping within the
metallic phase and become trapped at low temperatures, maybe by
the disorder potential associated with the existence of two
different orientations for the C$_{60}$ molecules in the $sc$
phase.

The most important consequence of this study is that it reveals an
\textit{attractive interaction at the local scale} which allows to
form spin-singlets despite the large Coulomb repulsion that should
forbid a double occupancy of the same site. We propose that this
interaction is mediated by Jahn-Teller distortions. Because the
gain of energy associated with a JTD is larger for an evenly
charged molecule, these configurations could be stabilized even in
compounds with odd stoichiometries. It is natural to wonder
whether this also plays a role in the case of A$_3$C$_{60}$.

The unexpected splitting of the $^{133}$Cs spectrum, which is the
consequence of the presence of C$_{60}^{2-}$, inevitably refers to
the T' line observed in many $fcc$ A$_3$C$_{60}$ systems for the
alkali
 in the tetrahedral site  \cite
{Pennington}. SEDOR experiments have shown that the T' line arises
from a modified tetrahedral site. Its intensity reproducibly
corresponds to about 15 \% of the tetrahedral sites. Many
explanations have been proposed for the origin of the T' site (T'
could be neighbor from
 a vacancy, a misoriented C$_{60}$ or a JTD one etc.) but none of them is
consistent with all experimental observations. By analogy to CQ
CsC$_{60}$ where the splitting is much clearer, it is tempting to
say that the T' line could correspond to a small fraction of about
3 \% localized C$_{60}^{2-}$ or C$_{60}^{4-}$ in A$_3$C$_{60}$,
maybe trapped by some defects. We have suggested that the $sc$
structure efficiently traps the C$_{60}^{2-}$ on the balls in the
minor orientation and the absence of such orientations in the
$fcc$ A$_3$C$_{60}$ compounds might play a role. To check this, it
would be interesting to know whether there is a similar line in
the orientationally ordered Na$_{2}$CsC$_{60}$ and what is its
intensity. This might be a clue to prove that the peculiar
properties of CQ CsC$_{60}$ actually reveal a common tendency in
alkali fullerides.

Indeed, an independent NMR observation, based on an anomalous
temperature dependence of 1/T$_1$T,
 has recently led us to suggest that C$%
_{60}^{2-} $ and C$_{60}^{4-}$ are also formed in A$_3$C$_{60}$
but usually on very short lifetimes of the order of 10$^{-14}$
sec. We have proposed this idea by studying Na$_2$CsC$_{60}$
\cite{BrouetPRL2001} and we will present in paper III a
generalization of this phenomenon to other systems like
Rb$_3$C$_{60}$. The main difference between CQ CsC$_{60}$ and
these other compounds then resides in the {\it lifetime} of the
C$_{60}^{2n-}$. In addition to the structural effect discussed
before, this could be related to the fact that neutral C$_{60}$ is
not stabilized by JTD, so that a jump of one electron is less
likely to lead to a stable configuration than
in A$_3$C$_{60}$, which has a symmetric position between C$_{60}^{2-}$ and C$%
_{60}^{4-}$. The different lifetime of the electronic pairs formed
via Jahn-Teller distortions might also explain why CQ CsC$_{60}$
is not superconducting contrary to A$_3$C$_{60}$. The difference
of behavior should vanish as one increases the temperature and the
singlets start to move more freely. Above 300 K, CsC$_{60}$ can be
studied again in a cubic structure, but, unexpectedly, it appears
to be insulating. We will show new data at high temperature in
paper III that allow to conclude that this behavior can be
understood by taking into account the presence of C$_{60}^{2-}$
and is in fact similar to that of some A$_3$C$_{60}$ systems.


\begin{thebibliography}{99}
\bibitem{pulse} To observe an NMR echo using the usual sequence
$\pi /2-\tau-\pi $, the length of the $\pi /2$ pulse is reduced if
only the central transition is observed. Furthermore, while the
length for $\pi $ should be exactly twice the $\pi /2$ one in the
latter case, it can be reduced to half this value when all the
transitions are observed, if the dephasing between the NMR spins
is dominated by a quadrupolar origin. In our case, the signal is
optimized for the two main lines by the same length of $\pi /2$
and a shorter value for $\pi $.
\end{thebibliography}
\end{document}